# Learning mixed graphical models from data with $p$ larger than $n$

**Inma Tur and Robert Castelo**
Dept. of Experimental and Health Sciences
Universitat Pompeu Fabra
Barcelona, Spain
{inma.tur, robert.castelo}@upf.edu

## Abstract

Structure learning of Gaussian graphical models is an extensively studied problem in the classical multivariate setting where the sample size $n$ is larger than the number of random variables $p$, as well as in the more challenging setting when $p \gg n$. However, analogous approaches for learning the structure of graphical models with mixed discrete and continuous variables when $p \gg n$ remain largely unexplored. Here we describe a statistical learning procedure for this problem based on limited-order correlations and assess its performance with synthetic and real data.

## 1 INTRODUCTION

Graphical Markov models (GMMs) are a powerful tool to work with families of multivariate distributions sharing a subset of conditional independence restrictions that can be represented by means of a graph. Different types of graph determine distinct classes of GMMs. For some of these classes, like those determined by undirected graphs and by acyclic digraphs (also known as DAGs or Bayesian networks), learning their structure from data is a well-studied problem in the classical framework where the sample size $n$ is much larger than the number of random variables (r.v.) $p$. In this setting, there are consistent structure estimation procedures (e.g., Lauritzen, 1996; Chickering, 2002; Castelo and Kočka, 2003) that converge to the generative structure in the limit of the size $n$ of the data under certain assumptions on the underlying probability distribution.

In the last decade, technological advances in the instrumentation employed in fields like physics, engineering or molecular biology have facilitated a continuous increase of the number of objects that these instruments simultaneously observe and quantify. Each such data set forms thus a multivariate sample of $n$ observations through a typically much larger number $p$ of r.v., i.e., where $p \gg n$. In this other setting, traditional assumptions underlying learning procedures do not hold and a substantial amount of work in GMM research has been devoted to the problem of learning the structure of the graph from data with $p \gg n$. Most of these contributions have been developed for Gaussian GMMs learned from pure continuous (multivariate normal) data, and they can be broadly categorized in regularization techniques (e.g., Friedman et al., 2008), dimension-reduction procedures (e.g., Segal et al., 2006) and limited-order correlations (e.g., Castelo and Roverato, 2006).

An important fraction of the current extraordinary growth of data sets with $p \gg n$ is produced in the biomedical field where high-throughput experimental technologies, coupled with comprehensive epidemiological and clinicopathological surveys, are profiling individuals at molecular, genotype, and clinical level. The simultaneous assay and recording of these data types generates multivariate samples of mixed discrete and continuous variables amenable for their analysis with mixed GMMs (Lauritzen and Wermuth, 1989).

However, most of the tools developed in molecular biology that deal with these data, rely on linear modeling techniques outside the mixed GMM framework (Broman and Sen, 2009), which has not been yet fully exploited. A recent contribution (Edwards et al., 2010; Abreu et al., 2010), employing mixed GMM theory to address this problem, provides a feasible and efficient solution by restricting the class of mixed GMMs to those that are decomposable.

In this paper we address the problem of learning the structure of non-decomposable mixed GMMs adapting a limited-order correlation approach previously introduced by Castelo and Roverato (2006) for Gaussian GMMs, and illustrate its feasibility and accuracy with both synthetic and real data. The rest of the paper is organized as follows. Section 2 reviews the necessary theory of mixed GMMs. In Section 3 we introduce limited-order correlations for mixed GMMs, and their performance for structure learning is assessed in Section 4 with both synthetic and real data. Finally, a short discussion is provided in Section 5.

## 2 MIXED GRAPHICAL MARKOV MODELS

In this section we review part of the mixed GMM theory that we need throughout the paper. Full details can be found in the books of Lauritzen (1996) and Edwards (2000). Mixed GMMs are GMMs for distributions involving discrete r.v., denoted by $I_\delta$ with $\delta \in \Delta$, and continuous r.v., denoted by $Y_\gamma$ with $\gamma \in \Gamma$, such that we define an undirected marked graph $G = (V, E)$ with $p$ marked vertices $V = \Delta \cup \Gamma$ , and edge set $E \subseteq V \times V$, where solid and open circles indicate vertices $\delta \in \Delta$ and $\gamma \in \Gamma$, respectively. An important subclass of these graphs is formed by decomposable marked graphs. When $G$ is undirected, decomposability holds if and only if $G$ does not contain chordless cycles of length larger than 3 and does not contain any path between two non-adjacent discrete vertices passing through continuous vertices only. In Figure 1 we show examples of such marked graphs.

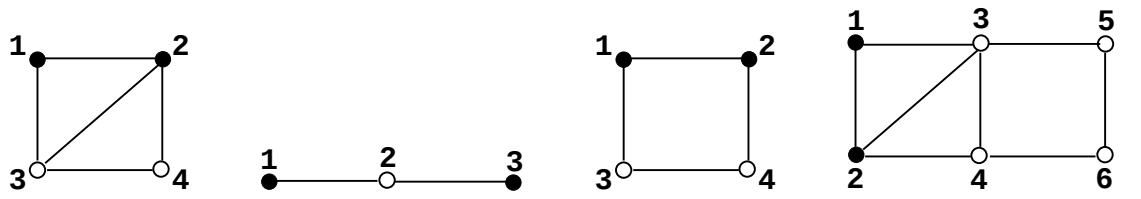

(a) (b) (c) (d)

Figure 1: Examples of graphs representing mixed GMMs: (a) is decomposable; (b), (c) and (d) are not.

The vertices of $G$ index a vector of r.v. $X = (I, Y)$ with elements of the joint space denoted by:

$$x = (i, y) = \{(i_\delta)_{\delta\in\Delta}, (y_\gamma)_{\gamma\in\Gamma}\} \ , \qquad (1)$$

where $i_\delta$ are discrete values and $y_\gamma$ are continuous. The set of all possible joint levels $i$ of the discrete variables is denoted by $\mathcal{I}$. Following (Lauritzen and Wermuth, 1989) we will assume that the joint distribution of the variables $X$ is conditional Gaussian (also known as CG-distribution) with density function:

$$f(x) = f(i, y) = p(i)|2\pi\Sigma(i)|^{-\frac{1}{2}} \times \exp\left\{-\frac{1}{2}(y - \mu(i))^T \Sigma(i)^{-1}(y - \mu(i))\right\} . \qquad (2)$$

This distribution has the property that continuous variables follow a multivariate normal distribution $\mathcal{N}_{|\Gamma|}(\mu(i), \Sigma(i))$ conditioned on the discrete variables. The parameters $(p(i), \mu(i), \Sigma(i))$ are called moment characteristics: $p(i)$ is the probability that $I = i$ and $\mu(i)$ and $\Sigma(i)$ are the conditional mean and the covariance matrix of $Y$ which may depend on $i$. If the covariance matrix is constant across the levels of $I$, that is, $\Sigma(i) \equiv \Sigma$, the model is *homogeneous.* Otherwise, the model is said to be *heterogeneous.* We can write the logarithm of the density in terms of the canonical parameters $(g(i), h(i), K(i))$:

$$\log f(i, y) = g(i) + h(i)^T y - \frac{1}{2} y^T K(i) y \ , \ \text{ where} \ , \qquad (3)$$

$$\begin{aligned} g(i) &= \log(p(i)) - \frac{1}{2}\log|\Sigma(i)| - \\ &\quad \frac{1}{2}\mu(i)^T\Sigma(i)^{-1}\mu(i) - \frac{|\Gamma|}{2}\log(2\pi) \ , \qquad (4) \\ h(i) &= \Sigma(i)^{-1}\mu(i) \ , \qquad (5) \\ K(i) &= \Sigma(i)^{-1} \ . \qquad (6) \end{aligned}$$

These terms can be expanded as follows:

$$g(i) = \sum_{d:d\subseteq\Delta} \lambda_d(i), \ h(i) = \sum_{d:d\subseteq\Delta} \eta_d(i), \ K(i) = \sum_{d:d\subseteq\Delta} \Psi_d(i) \ . \qquad (7)$$

Plugging these expansions in equation (3) we obtain

$$\log f(i, y) = \sum_{d\subseteq\Delta} \lambda_d(i) + \sum_{d\subseteq\Delta}\sum_{\gamma\in\Gamma} \eta_d(i)_\gamma y_\gamma - \frac{1}{2}\sum_{d\subseteq\Delta}\sum_{\gamma,\eta\in\Gamma} \psi_d(i)_{\gamma\eta} y_\gamma y_\eta \ , \qquad (8)$$

where the interaction terms are described as:

- $\lambda_d(i)$, with $d \neq \emptyset$ are the discrete interactions among the variables indexed by $d$. If $|d| = 1$, the term is called main effect of the variable in $d$. If $d = \emptyset$ the term $\lambda_\emptyset$ is constant.
- $\eta_d(i)_\gamma$, with $d \neq \emptyset$, represent mixed linear interactions between $X_\gamma$ and the variables indexed by $d$. If $d = \emptyset$, the term $\eta_{\emptyset\gamma}$ is called linear main effect of the variable $X_\gamma$.
- $\psi_d(i)_{\gamma\eta}$ represent quadratic interactions between $X_\gamma, X_\eta$ and the variables indexed by $d$. If $\gamma = \eta$ and $d = \emptyset$, we speak of quadratic main effects. If the model is homogeneous, there are not mixed quadratic interactions, i.e., $\Psi_d = 0$ for $d \neq \emptyset$.

Let $\{x^{(\nu)}\} = \{(i^{(\nu)}, y^{(\nu)})\}$ be a sample of $\nu = 1, ..., n$ independent and identically distributed observations from a CG-distribution. For an arbitrary subset $A \subseteq V$, we abbreviate to $i_A = i_{A\cap\Delta}$, $\mathcal{I}_A = \mathcal{I}_{\Delta\cap A}$ and $y_A = y_{A\cap\Gamma}$ and the following sampling statistics are defined:

- $n(i) = \#\left\{\nu : i^{(\nu)} = i\right\}$ = nr. level-$i$ observations.
- $s(i) = \sum_{\nu:i^{(\nu)}=i} y^{(\nu)}$ = sum of the level-$i$ $y$-values.
- $\bar{y}(i) = s(i)/n(i)$ = sample mean of level-$i$ $y$-values.
- $ss(i) = \sum_{\nu:i^{(\nu)}=i} y^{(\nu)}(y^{(\nu)})^T$ = sum of squares of level-$i$ $y$-values.
- $ssd(i) = ss(i) - s(i)s(i)^T/n(i)$ = sum of squares of deviations from the mean of level $i$.
- $ssd(A) = \sum_{i_A \in \mathcal{I}_A} ssd(i_A)$ = sum of squares of deviations from the mean within the cells of $\mathcal{I}_A$.
- $ssd_A(A) = \sum_{i_A \in \mathcal{I}_A} ssd_{A\cap\Gamma}(i_A)$.
- $ssd = ssd(V)$ and $ssd_A = ssd_A(V)$.

Lauritzen (1996, Prop. 6.9) shows that the likelihood function for the heterogeneous, saturated model attains its maximum if and only if $ssd(i)$ is positive definite for all $i \in \mathcal{I}$, which is almost surely equal to the event that $n(i) > |\Gamma|$ for all $i \in \mathcal{I}$. If the maximum likelihood estimate exists, it is given as having moment characteristics equal to the empirical moments, i.e.,

$$\hat{p} = n(i)/n, \quad \hat{\mu}(i) = \bar{y}(i), \quad \hat{\Sigma}(i) = ssd(i)/n(i)\,. \quad (9)$$

This sample size constraint is milder with the homogeneous, saturated model whose likelihood function attains its maximum if and only if $n(i) > 0$ for all $i \in \mathcal{I}$ and $ssd$ is positive definite (Lauritzen, 1996, Prop. 6.10), which cannot occur whenever $n < |\Gamma| + |\mathcal{I}|$. If $n \geq |\Gamma| + |\mathcal{I}|$ then this is almost surely equal to the event that $n(i) > 0$ for all $i \in \mathcal{I}$. In such a case, the moment characteristics are the same as in equation (9) for the heterogenous model, except for the covariance matrix which is constant across the levels of the discrete variables and, therefore, $\hat{\Sigma} = ssd/n$.

From these latter results it follows that mixed GMMs cannot be directly estimated from data with $p \gg n$, using only the formulae described in this section.

# 3 LIMITED-ORDER CORRELATIONS

We address the problem of learning mixed GMMs from data with $p \gg n$ by using a limited-order correlation approach, previously introduced for Gaussian GMMs by Castelo and Roverato (2006). The fundamental idea consists of estimating, for every pair of r.v., a linear measure of association over all marginal distributions of size $(q + 2) < n$. An approximation to the underlying graph $G$, called qp-graph and denoted by $G^{(q)}$, can be obtained by removing edges from a complete graph, that do not meet a given cutoff on this measure. In (Castelo and Roverato, 2006) such a measure of association is introduced with the, so-called, *non-rejection rate*, as follows.

Let $\mathcal{Q}^q_{\alpha\beta} = \{Q \subseteq V\backslash\{\alpha,\beta\} : |Q| = q\}$. Let $T^q_{\alpha\beta}$ be a binary random variable associated to the pair of vertices $(\alpha,\beta)$ that takes values from the following three-step procedure: 1. an element $Q$ is sampled from $\mathcal{Q}^q_{\alpha\beta}$ according to a (discrete) uniform distribution; 2. test the null hypothesis of conditional independence $H_0 : X_\alpha \perp\!\!\!\perp X_\beta | X_Q$ for which, in the Gaussian GMM case, Castelo and Roverato (2006) employed a test of zero regression coefficient whose statistic under the null follows a $t$-student distribution exactly; and 3. if the null hypothesis $H_0$ is rejected then $T^q_{\alpha\beta}$ takes value 0, otherwise takes value 1. It follows that $T^q_{\alpha\beta}$ has a Bernoulli distribution and the non-rejection rate is defined as its expectancy $\mathrm{E}[T^q_{\alpha\beta}] = Pr(T^q_{\alpha\beta} = 1)$. Since the number of subsets $Q$ in $\mathcal{Q}^q_{\alpha\beta}$ can be very large, the authors in (Castelo and Roverato, 2006) propose to uniformly sample only a limited number of them like, for instance, one-hundred.

Therefore, applying such an strategy to mixed continuous and discrete data basically amounts to find a suitable test for $H_0 : X_\alpha \perp\!\!\!\perp X_\beta | X_Q$. However, this approach requires performing many tests with often small sample sizes $n$ and/or $n(i)$, since $p \gg n$. A way to address this consists of restricting these tests to models on the corresponding $(q+2)$ r.v. where the saturated $H_1$ and the constrained $H_0$ are decomposable, such that explicit maximum likelihood estimates exist (Lauritzen, 1996, pg. 188) and where the distribution of the statistic under the null is exact, which is preferred to an asymptotic one for small sample sizes.

In order to meet these requirements, we restrict the learning problem to mixed GMMs where discrete r.v. are marginally independent and thus their associations will not be considered. A further simplifying assumption is that we restrict the learning problem to homogeneous mixed GMMs, where the sample size requirements for the existence of maximum likelihood estimates are easier to meet with data where $p \gg n$.

## 3.1 ASYMPTOTIC AND EXACT CONDITIONAL INDEPENDENCE TESTS

Following (Lauritzen, 1996) we will test the conditional independence $X_\alpha \perp\!\!\!\perp X_\beta | X_Q$ by performing a likelihood-ratio test between two decomposable models: the saturated model determined by the complete graph $G^1 = (V, E^1)$ where $V = \{\alpha, \beta, Q\}$ and $E^1 = V \times V$, and the constrained model determined by $G^0 = (V, E^0)$ with exactly one missing edge formed by the two vertices $\alpha, \beta$ representing the r.v. we want to test, and thus $E^0 = \{V \times V\} \backslash (\alpha, \beta)$ and $Q = V \backslash \{\alpha, \beta\}$.

With $V = \Delta \cup \Gamma$, we denote by $(\gamma, \eta)$ a pair of continuous variables (i.e., $\gamma, \eta \in \Gamma$), by $(\delta, \gamma)$ a pair of mixed variables with $\delta \in \Delta$ and, again, $\gamma \in \Gamma$, so that either $Q = V \backslash \{\gamma, \eta\}$ or $Q = V \backslash \{\delta, \gamma\}$ as the conditioning subset. In the pure continuous case, the conditional independence $\gamma \perp\!\!\!\perp \eta | Q$ in an homogeneous mixed GMM corresponds to a zero value in the canonical parameter $K$ and the corresponding likelihood ratio statistic raised to the power $2/n$ is (Lauritzen, 1996, pg. 192):

$$\Lambda_{\gamma\eta.Q} = \frac{|ssd_\Gamma||ssd_{\Gamma \backslash \{\gamma,\eta\}}|}{|ssd_{\Gamma \backslash \{\gamma\}}||ssd_{\Gamma \backslash \{\eta\}}|} \,. \tag{10}$$

In the mixed case, the conditional independence $\delta \perp\!\!\!\perp \gamma | Q$ corresponds to an expansion of the canonical parameter $h(i)$ where the terms corresponding to $\delta$ are zero. The likelihood ratio statistic raised to the power $2/n$ is (Lauritzen, 1996, pg. 194):

$$\Lambda_{\delta\gamma.Q} = \frac{|ssd_\Gamma||ssd_{\Gamma^*}(\Delta^*)|}{|ssd_{\Gamma^*}||ssd_\Gamma(\Delta^*)|} \,, \tag{11}$$

where $\Gamma^* = \Gamma \backslash \{\gamma\}$ and $\Delta^* = \Delta \backslash \{\delta\}$. We will see below that the values of $-n \log \Lambda_{\gamma\eta.Q}$, when testing the presence of a pure continuous edge, and the values of $-n \log \Lambda_{\delta\gamma.Q}$ when testing the presence of a mixed edge, follow asymptotically a $\chi^2_{df}$ distribution with $df = 1$ and $df = |\mathcal{I}_{\Delta^*}|(|\mathcal{I}_\delta| - 1)$ degrees of freedom in the continuous and mixed case, respectively.

However, (Lauritzen, 1996, pg. 192 to 194) observes that for decomposable heterogeneous mixed GMMs, the likelihood ratio in equations (10) and (11) follows exactly a beta distribution with certain parameters. In order to enable the analogous exact test for homogeneous mixed GMMs, we proceed to derive their corresponding parameters. To that end, we will consider the fact that the joint distribution of $X_V$ is equivalent to the conditional one of $X_\gamma$ given the rest of the variables $X_{V \backslash \{\gamma\}}$. This equivalence exists due to the decomposability of the complete and the constrained models, which ensures the collapsibility of both models onto the same set of variables $X_{V \backslash \{\gamma\}}$ (see Edwards, 2000, Sec. 4.2).

Starting with the pure continuous case, the conditional expectation of $X_\gamma$ given the rest of the variables under the saturated model can be written as,

$$\mathrm{E}\left(X_\gamma | \Delta, \Gamma \backslash \{\gamma\}\right) = \alpha(i_\Delta) + \sum_{\lambda \in \Gamma \backslash \{\gamma\}} \beta_{\gamma\lambda|\Gamma \backslash \{\gamma\}} X_\lambda \,, \tag{12}$$

where $\alpha(i_\Delta) = \mu_\gamma(i_\Delta) - \sum_{\lambda \in \Gamma \backslash \{\gamma\}} \beta_{\gamma\lambda|\Gamma \backslash \{\gamma\}} \mu_\lambda(i_\Delta)$ and $\beta_{\gamma\lambda|\Gamma \backslash \{\gamma\}}$ is the partial regression coefficient that is found through the canonical parameter $K = \{k_{\gamma\eta}\}, \forall \gamma, \eta \in \Gamma$, as $\beta_{\gamma\lambda|\Gamma \backslash \{\gamma\}} = -k_{\gamma\lambda}/k_{\gamma\gamma}$ (Lauritzen, 1996, pg. 130). Since the first term in equation (12) involves $|\mathcal{I}|$ parameters and the second $|\Gamma| - 1$, the total number of degrees of freedom of the sum of squares of deviations of this model is $n - |\Gamma| - |\mathcal{I}| + 1$. Under the constrained model, the conditional expectation is,

$$\mathrm{E}\left(X_\gamma | \Delta, \Gamma \backslash \{\gamma, \eta\}\right) = \alpha(i_\Delta) + \sum_{\lambda \in \Gamma \backslash \{\gamma,\eta\}} \beta_{\gamma\lambda|\Gamma \backslash \{\gamma,\eta\}} X_\lambda, \tag{13}$$

which, in an analogous way to the saturated model, leads to $n - |\Gamma| - |\mathcal{I}| + 2$ degrees of freedom.

Let $RSS_1$ and $RSS_0$ denote the $\chi^2_k$-distributed residual sum of squares of $X_\gamma$ under the saturated and constrained models, respectively, and $RSS_{1.0}$, the difference $RSS_0 - RSS_1$. Given that a r.v. $X$ following a $\chi^2_k$ with $k$ degrees of freedom also follows a gamma distribution (Rao, 1973, pg. 166) with $\Gamma(k/2, 2)$, then,

$$RSS_1 \sim \Gamma\left(\frac{n - |\Gamma| - |\mathcal{I}| + 1}{2}, 2\right) \,, \tag{14}$$

$$RSS_0 \sim \Gamma\left(\frac{n - |\Gamma| - |\mathcal{I}| + 2}{2}, 2\right) \,, \tag{15}$$

and, thus, $RSS_{1.0} \sim \Gamma\left(1/2, 2\right)$. Moreover, if $X$ and $Y$ are two independent r.v. such that $X \sim \Gamma\left(k_1, \theta\right)$ and $Y \sim \Gamma\left(k_2, \theta\right)$, then (Rao, 1973, pg. 165),

$$\frac{X}{X + Y} \sim \mathcal{B}(k_1, k_2) \,, \tag{16}$$

where $\mathcal{B}(k_1, k_2)$ denotes the beta distribution with shape parameters $k_1$ and $k_2$. If we let $X = RSS_1$ and $Y = RSS_{1.0}$, then the ratio $RSS_1/RSS_0$, under collapsability (Edwards, 2000, pg. 87), is the same quantity as the ratio in (10) and we finally obtain that,

$$\Lambda_{\gamma\eta.Q} \sim \mathcal{B}\left(\frac{n - |\Gamma| - |\mathcal{I}| + 1}{2}, \frac{1}{2}\right) \,. \tag{17}$$

In the mixed case, the conditional expectation of $\gamma \in \Gamma$ given the rest of variables under the saturated model coincides with the continuous case in equation (12) and is slightly different under the constrained model,

$$\mathrm{E}\left(X_\gamma | \Delta \backslash \{\delta\}, \Gamma \backslash \{\gamma\}\right) = \alpha(i_{\Delta \backslash \{\delta\}}) + \sum_{\lambda \in \Gamma \backslash \{\gamma\}} \beta_{\gamma\lambda|\Gamma \backslash \{\gamma\}} X_\lambda \,. \tag{18}$$

The first term involves $|\mathcal{I}_{\Delta^*}|$ parameters and the second $|\Gamma| - 1$, so that the constrained model has $n - |\Gamma| - |\mathcal{I}_{\Delta^*}| + 1$ degrees of freedom. By an argument analogous to the pure continuous case, the likelihood ratio statistic raised to the power $2/n$ for the null hypothesis of a missing mixed edge follows a beta distribution with these parameters:

$$\Lambda_{\delta\gamma.Q} \sim \mathcal{B}\left(\frac{n - |\Gamma| - |\mathcal{I}| + 1}{2}, \frac{|\mathcal{I}_{\Delta^*}|(|\mathcal{I}_\delta| - 1)}{2}\right) . \quad (19)$$

# 4 RESULTS

In this section we show experimental results with synthetic and real data. First, we describe how we have generated synthetic data from mixed GMMs. Second, we verify that the exact conditional test for mixed data that we described in Section 3 controls the significance level across decreasing sample sizes and decreasing degrees of sparseness of the generative graph. Third, an exhaustive assessment of the accuracy of the method, in terms of precision-recall curves, is provided, in comparison with the method of Abreu et al. (2010). Fourth, its performance on real expression and genotype data from yeast is described.

## 4.1 SYNTHETIC HOMOGENEOUS MIXED GRAPHICAL MARKOV MODELS

We build synthetic homogeneous mixed GMMs by first sampling a graph structure $G = (V, E)$, specifying $\Delta$ and $\Gamma$ such that $V = \Delta \cup \Gamma$, and then by generating random parameters that convey the conditional independences encoded in $G$. In data with $p \gg n$, the sparseness of the graph structure that we want to learn has a direct impact in performance since the basic assumption of every learning approach in this setting is that the underlying structure is sparse. In order to have a fine-tune control on the sparseness of the graphs employed to build synthetic mixed GMMs we consider sampling graphs from the subclass of undirected $d$-regular graphs (Harary, 1969). A $d$-regular graph has a constant vertex degree $d$ for all its vertices, which bounds the size of any minimal subset separating every pair of vertices (Castelo and Roverato, 2006, pg. 2646) and its graph density is a linear function of $d$. We have used the algorithm by Steger and Wormald (1999) to sample $d$-regular graphs uniformly at random.

Since we assume that discrete r.v. are marginally independent between them, if two discrete vertices are connected in a sampled $d$-regular graph, the graph is rejected and a new graph is sampled till one is obtained where every pair of discrete vertices is disconnected.

Once the structure of a $d$-regular graph $G$ is obtained, a random covariance matrix $\Sigma$ is generated with unit diagonal, off-diagonal elements corresponding to marginal Pearson correlations with a mean value $\rho$ that we specify, and whose pattern of zeros in its inverse matches the missing edges in $G$. In order to guarantee the positive definiteness of $\Sigma$, the specified value $\rho$ should be such that $-1/(p-1) < \rho < 1$.

The random vector $\mu(i)$, conditional on the discrete levels $\mathcal{I}$, is generated using equation (5) so that:

$$\mu(i) = \Sigma \cdot h(i) . \quad (20)$$

The values of canonical parameters $h(i) = \{h_\gamma(i)\}, \gamma \in \Gamma$, determine the strength of the mixed linear interactions between discrete and continuous r.v. They are generated by setting, for each $\gamma \in \Gamma$, $h_\gamma(i) = \{z_{\gamma i_A}\}_{i_A \in \mathcal{I}}$, where $A \subseteq \Delta$ and every $\delta \in A$ forms an edge with $\gamma$, i.e., $(\gamma, \delta) \in E$. Values $z_{\gamma i_A}$ are sampled from a normal distribution $\mathcal{N}(0, \sigma)$ where $\sigma$ determines the magnitude in which values of $h_\gamma(i)$ and $h_\gamma(j)$, $i \neq j$, differ throughout each $i \in \mathcal{I}$. This, in turn, determines the strength of the linear mixed interaction with larger $\sigma$ values leading to stronger mixed interactions. For every $\gamma \in \Gamma$ with no interaction with any discrete variable one single value $z_\gamma \sim \mathcal{N}(0, \sigma)$ is sampled and assigned to $h_\gamma(i)$ for every $i \in \mathcal{I}$.

In order to make the discrete r.v. $X_\Delta$ marginally independent between them, their joint levels are assigned with a uniform distribution. Finally, every observation is generated by first sampling a joint level $i \in \mathcal{I}$ of $X_\Delta$ according to their (uniform) probability distribution, and, secondly, by sampling a multivariate normal observation from $\mathcal{N}(\mu(i), \Sigma)$ for the continuous r.v. $X_\Gamma$.

## 4.2 CONDITIONAL INDEPENDENCE TEST ERROR CONTROL

We want to verify that the exact conditional independence test described in section 3 provides an accurate control of its significance level. We consider two null hypotheses to test, a missing continuous and a missing mixed edge on four vertices, where two are discrete and two are continuous. In order to satisfy the assumption of marginal independence between the discrete r.v., the joint distributions of these two synthetic models are represented in Figure 2 by chain graphs (Edwards, 2000, Sec. 7.2). Using these graphs and the procedure described before, we generate two sets of parameters.

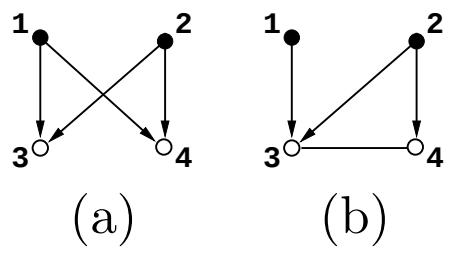

Figure 2: Chain graphs with one missing continuous edge in (a) and one missing mixed edge in (b).

From each of these two synthetic models we have sampled 10,000 data sets of sizes 100, 75, 50 and 25, and from each of them, the corresponding likelihood ratio for the missing continuous and missing mixed edge, was calculated. Using the theoretical quantile at $\alpha = 0.05$, we estimated the Type-I error probability from the ranking of 10,000 likelihood ratios, calculated for the asymptotic and the exact test, and plot the results in Figure 3a. As expected, the exact test provides a better control of the significance level than the asymptotic one keeping it approximately constant through decreasing sample sizes. We have examined also the case in which sample size is fixed at $n = 25$ and graph density increases by considering $p = 50$ r.v., where 2 are discrete and 48 are continuous, and $d$-regular graphs are sampled with 5 different constant degrees from 3 to 7. Panel (b) of Figure 3 shows the empirical Type-I error probability as function of the graph density for two arbitrarily chosen missing, continuous and mixed, edges. We also observe that the exact test provides a better control of the probability of a Type-I error as sparseness decreases.

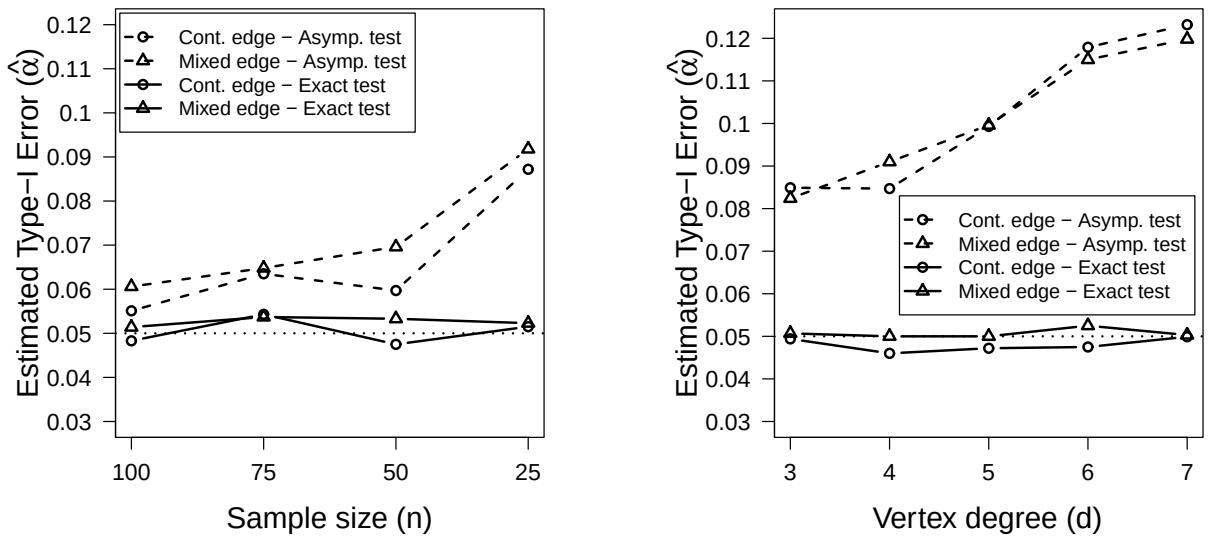

(a) (b)

Figure 3: Comparison of the estimated Type-I error $\hat{\alpha}$ between asymptotic and exact tests at a nominal level $\alpha = 0.05$ (dotted horizontal line) for missing continuous and mixed edges: (a) as function of the sample size $n$; (b), given $n = 25$, as function of vertex degree.

### 4.3 PERFORMANCE WITH SYNTHETIC DATA

We want to assess now the performance of the non-rejection rate for homogeneous mixed GMMs with synthetic data. We consider $d$-regular graphs of $p = 50$ vertices, where 2 of them correspond to discrete variables, 48 to continuous ones, and 3 different vertex degrees $d = \{3, 4, 7\}$. Four increasing values of nominal mean Pearson correlations $\rho = \{0.2, 0.4, 0.6, 0.8\}$ are employed to generate random covariance matrices $\Sigma$ whose inverse $K = \Sigma^{-1}$ has a zero pattern on the missing edges between the continuous r.v. $X_\Gamma$. Analogously, four increasing standard deviation values $\sigma = \{1, 2, 3, 4\}$ are set for sampling mixed linear interaction parameters $\{h(i)\}$, in correspondence with the values in $\rho$ in order to simultaneously increase the strength of both continuous and mixed edges. Five such covariance matrices and mixed linear interaction parameters are sampled per graph and, with them, corresponding mean vectors $\mu(i)$ of the CG-distributions are generated.

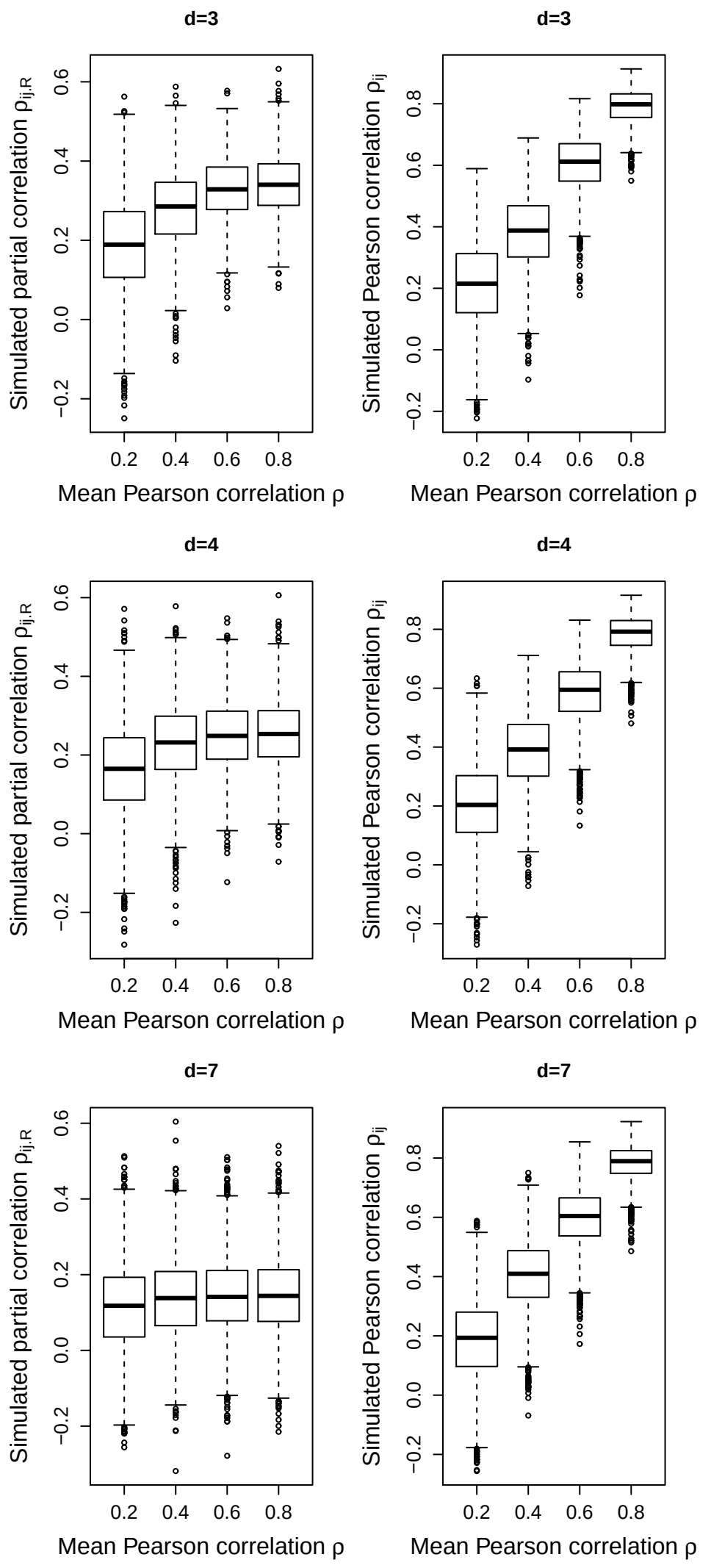

Figure 4: Simulated values of partial and Pearson correlation coefficients as function of the nominal mean Pearson correlation.

The values of the simulated Pearson and partial correlations, resulting from the sampled covariance matrices, are displayed in Figure 4 as function of the nominal mean Pearson correlation. As in the covariance decomposition model proposed by Jones and West (2005), partial correlations approach zero as the density of the graph increases. In short, the simulated parameters cover a wide spectrum of homogeneous mixed GMMs from which synthetic data can be generated.

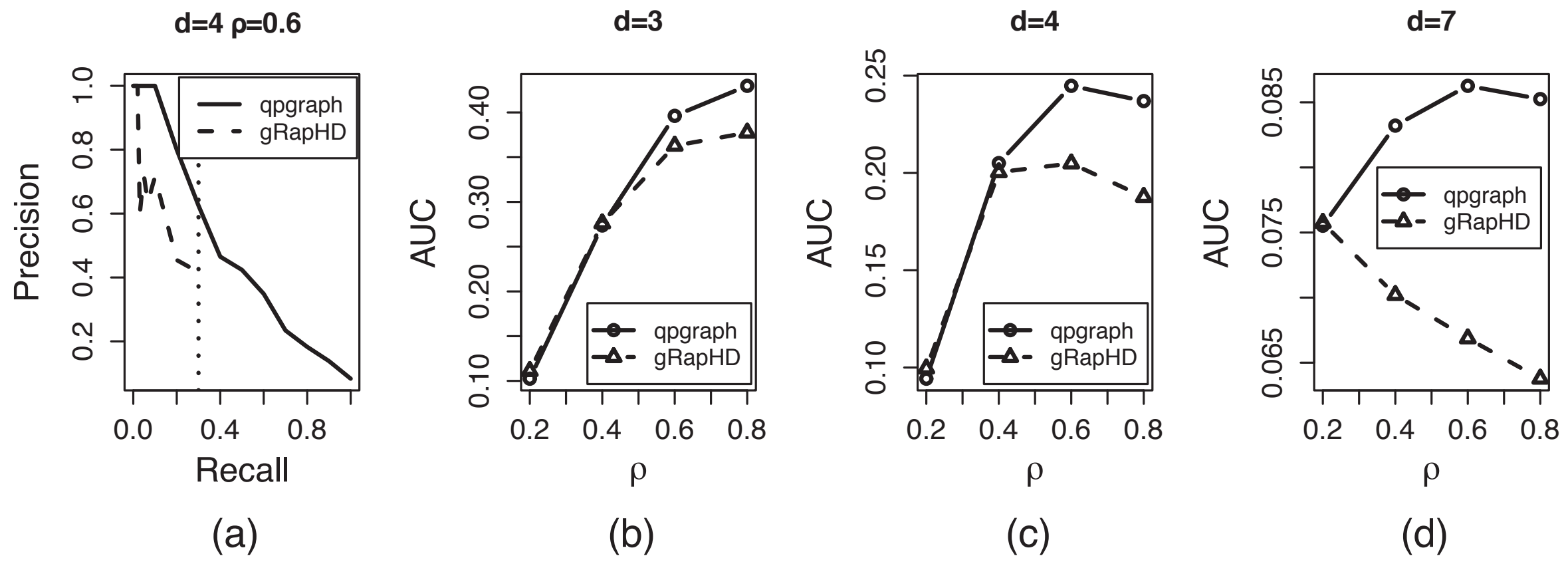

Figure 5: Performance with synthetic data. In panel (a) two specific precision-recalls are drawn from one of the data sets simulated from a regular graph with $d = 4$ and covariance matrix with $\rho = 0.6$. The vertical dot bar indicates the recall level attained by the gRapHD method. In panels (b, c, d) the values of the areas under the curve (AUC) are shown for increasing graph density and as function of the nominal mean Pearson correlation.

For each set of parameters, five data sets are sampled of $n = 25$ observations and from each data set, non-rejection rates are estimated using a small order of the correlations $q = 3$. The approach in (Abreu et al., 2010), implemented in the R package `gRapHD`, has been also assessed with these data and we report their performance in terms of precision-recall curves with respect to the generating graph. However, since `gRapHD` searches for decomposable models using a stepwise forward selection approach that adds at each step the edge that preserves decomposability and minimizes the BIC criterion, we have restricted the precision-recall curve comparison to the recall level attained by `gRapHD`, as illustrated in panel (a) of Figure 5. In that figure our method is denoted by the term "qpgraph".

On each data set, the area under the precision-recall curve (AUC), bounded by the maximum recall of `gRapHD`, was calculated. Panels (b, c, d) of Figure 5 show the average AUC value per parameter strength value across the combination of 5 graphs, 5 parameter sets and 5 data sets (i.e., across 125 points). As Figure 5 shows, both methods decrease their performance in such a $p \gg n$ setting as the complexity of the graph increases. However, this happens more dramatically when restricting the search space to decomposable graphs, as the `gRapHD` approach does.

### 4.4 PERFORMANCE WITH REAL DATA

Here we assess the performance of the method in a real data set from a study by Brem and Kruglyak (2005) where two yeast strains, a wild-type and a lab strain, were crossed to generate 112 segregants which were profiled in their gene expression and genotyped. The resulting data consist of 6,216 genes and 2,906 genotype markers throughout 112 samples. We have performed a simple expression Quantitative Trait Loci (eQTL) analysis by single marker regression using the `qtl` package (Broman and Sen, 2009) where missing genotypes have been previously imputed using the hidden Markov model approach implemented in the `sim.geno()` function of this package.

In Figure 6 we show an incidence matrix of all marker-gene pairs where black cells indicate that the LOD score for that particular marker-gene association is significant with a P-value $< 0.01$ according to the permutation test employed by `qtl`. In this matrix, genes and markers are ordered according to their position along the genome where chromosomes, indicated by roman numerals, are also arranged in increasing order. The diagonal pattern shows *cis*-acting associations, which correspond to genetic variation affecting the expression of the gene occurring close to where the marker is located. Off-diagonal associations correspond to *trans*-acting effects where genetic variation is, in principle, affecting the expression of genes located in other loci in the genome (see Rockman, 2008, for a detailed description of these concepts). Vertical bands correspond to loci in the genome whose genetic variation affects the expression of a large number of genes. These loci are known as *eQTL hotspots* (Breitling et al., 2008) and an important question is what fraction of this large number of affected genes are directly associated to the marker, and what other fraction does it indirectly.

In principle, a multivariate approach such as the one presented in this paper, should do better than an uni-variate one at distinguishing direct from indirect associations. In order to assess this hypothesis we have considered the 4 eQTL hotspots from different chromosomes associated to the largest number of genes (> 150), which are indicated by arrows in Figure 6. Using the data described before we have estimated non-rejection rates (NRR) for every of the four markers and every gene expression profile, restricting $\mathcal{Q}^q_{\alpha\beta}$ to subsets in $\Gamma \backslash \{\alpha, \beta\}$. Since the sample size is $n = 112$, we have employed different values of $q$ spanning the available range, more concretely $q = \{25, 50, 75, 100\}$, and then we have taken the average of the rates for each marker-gene pair. Castelo and Roverato (2009) showed that averaging the non-rejection rate is a sensible strategy to avoid having to choose a single $q$ value.

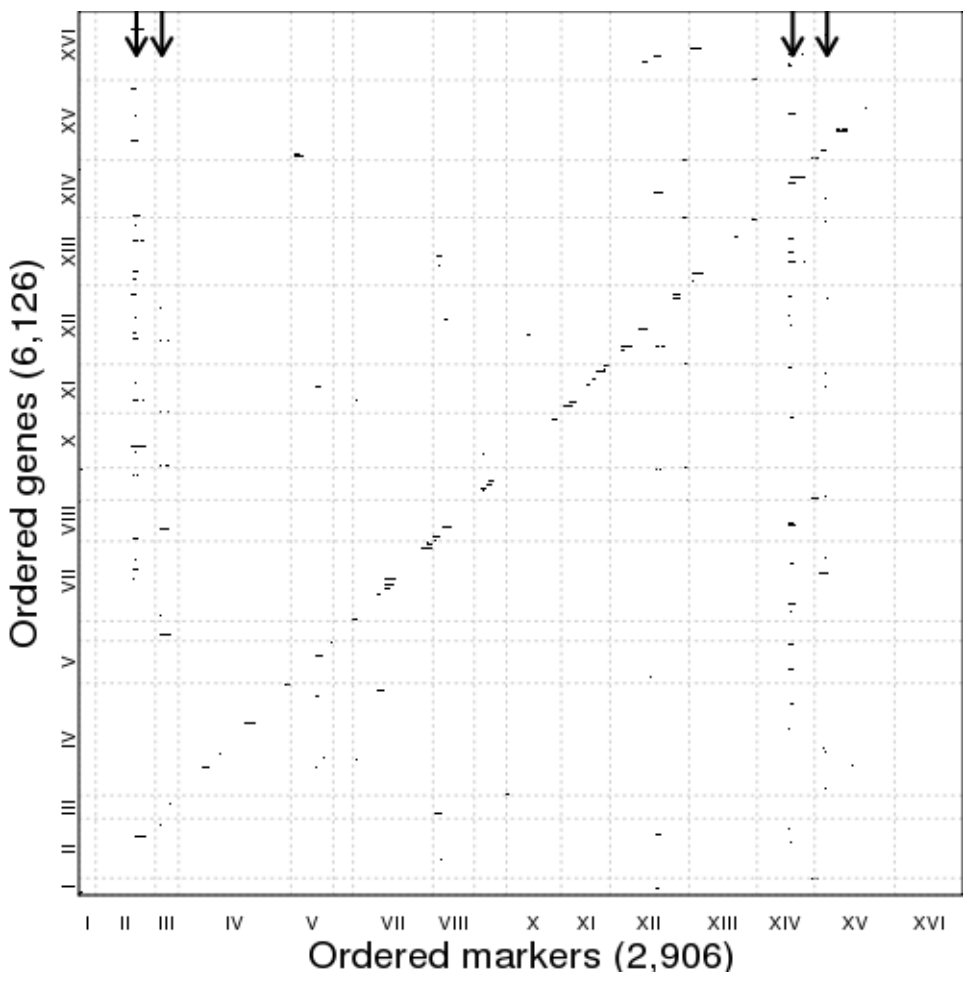

Figure 6: Significant associations found by single marker regression. Arrows indicate hotspot markers affecting the expression levels of more than 150 genes.

Using NRR values and the LOD scores of single marker regression we built two rankings of marker and gene expression associations. We observed that no association for the hotspot in chromosome 14 ranked among the first 40 and 25 associations for NRR and LOD, respectively, and discarded this hotspot from further analysis. Then, we took the top-$k$ genes in each of the two rankings for different values of $k = \{20, 30, 40, 50, 60, 70\}$ and for each hotspot we estimated the degree of functional coherence (FC) in an analogous way to the work of Castelo and Roverato (2009), where this approach was used for transcriptional networks. For each hotspot, we first retrieved the functional annotation[1] of the genes at less than 1kb. Second, among those genes that form part of the considered top-$k$ fraction, we calculated functionally enriched GO categories by a conditional hypergeometric test (Falcon and Gentleman, 2007). Finally, FC was estimated as the overlap in the GO hierarchies above the two sets of functional annotations, where higher values indicate larger coincidence between the molecular function exerted by the gene proximal to the hotspot and the genes at the top of the ranking, thus more directly connected to the marker. In Figure 7, panel (a), we show the distribution of FC values across each of the 6 top-$k$ gene subsets. The method presented in this paper provides higher mean and median values of FC. This implies that in the underlying unknown molecular network, genes affected in *cis* by the hotspot are in some functional sense "closer" to the genes that our method puts on the top of the ranking. One such example is the hotspot in chromosome 3 where the top 3 genes connected by NRR values and LOD scores are shown in panel (b). According to the annotations at the UCSC Genome Browser[2], and shown in panel (c), the strongest association by NRR is *cis*-acting with the marker and the other two are downstream of a binding site of LEU3. This gene is a major regulatory switch in the pathway of LEU2, whose activity may be affected by a feedback loop in the pathway (Chin et al., 2008).

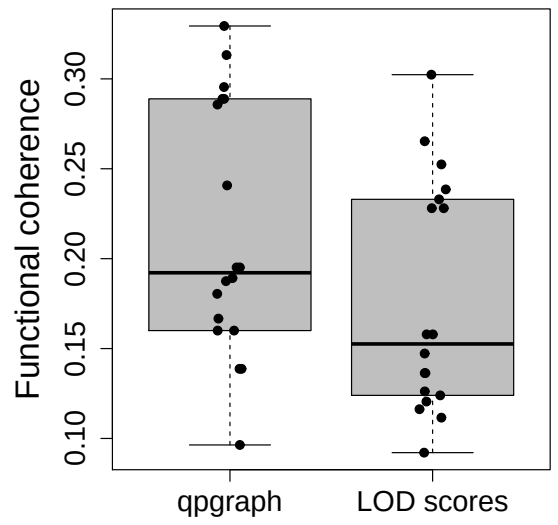

(a)

**Hotspot marker at chr3:90790**

| Gene | NRR |
|---|---|
| LEU2 | 0.02 |
| LEU1 | 0.06 |
| OAC1 | 0.08 |

| Gene | LOD |
|---|---|
| RLF2 | 40.58 |
| BZZ1 | 32.04 |
| PHR1 | 25.47 |

(b)

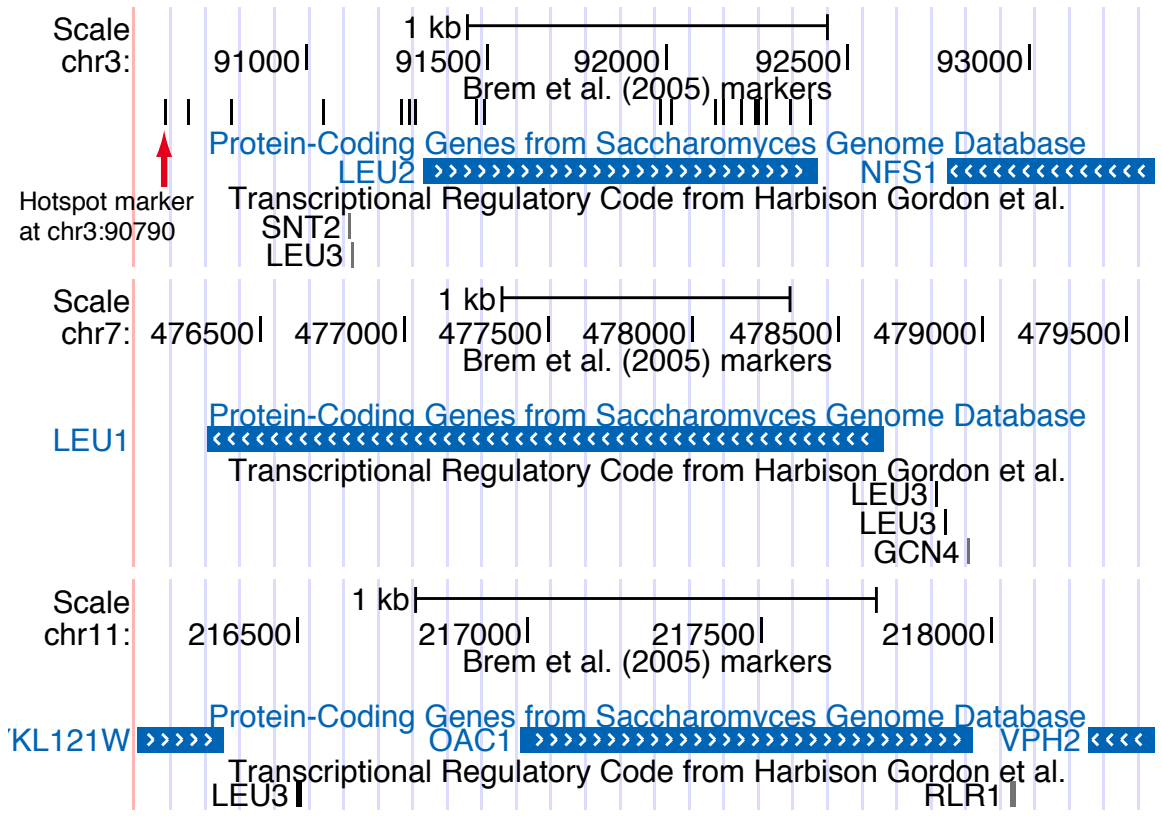

(c)

Figure 7: Yeast data analysis: (a) FC values between genes proximal to the hotspot markers and the top-$k$ ranked genes; (b) top 3 associations for the hotspot at chromosome 3 and their genomic context in (c).

[1]Using the GO database http://www.geneontology.org

[2]http://genome.ucsc.edu

# 5 DISCUSSION

In this paper we have introduced a statistical procedure to learn GMMs from mixed continuous and discrete data with $p \gg n$. We have adapted the limited-correlation approach of Castelo and Roverato (2006) for Gaussian GMMs to mixed GMMs by using an exact test of conditional independence, which we have shown, through simulation, to be suitable for testing with small sample sizes. We have also investigated the performance of the method as function of the graph density and of the strength of the correlations through a total of 1,500 simulated data sets. The analysis with real molecular data from yeast also showed that this approach may help in finding more direct associations between genetic variation and gene expression. In summary, limited-order correlations and marginal distributions constitute an appealing framework to develop approaches that exploit the sparseness of the underlying network when trying to learn the structure of a mixed GMM from data with $p \gg n$. The methodology presented in this paper is implemented through the function `qpNrr()` that forms part of the Bioconductor package `qpgraph` available from http://www.bioconductor.org.

**Acknowledgements**

We thank the anonymous reviewers for their remarks. We also thank David Cox, Steffen Lauritzen and Alberto Roverato for helpful discussions on several aspects of the paper. The first author is supported by a FPI predoctoral fellowship [BES-2009-024901] and the second author is a research fellow from the "Ramon y Cajal" program [RYC-2006-000932], both funded by the Spanish Ministerio de Ciencia e Innovación (MICINN). This work is supported by a MICINN project grant [TIN2008-00556/TIN] and part of it was developed while the first author was visiting the Dept. of Statistics at Oxford University, funded by a short-term visit MICINN fellowship [EEBB-2011-43932].